\documentclass[%
 reprint,
 superscriptaddress,
 amsmath,amssymb,
 aps,
 prb,
]{revtex4-2}
\usepackage{graphicx}
\graphicspath{ {./images/} }
\usepackage[utf8]{inputenc}
\usepackage{xcolor}
\usepackage{hyperref}
\usepackage[strings]{underscore}
\usepackage{bm}
\usepackage{ulem}

\begin{document}

 \title{Strain-tuned topological phase transition and unconventional Zeeman effect in  \texorpdfstring {ZrTe$_5$}{lg} microcrystals}

\author{Apurva Gaikwad} 
\thanks{Denotes equal contribution}
\affiliation{Department of Physics and Astronomy, Stony Brook University, Stony Brook NY 11794, USA}
\author{Song Sun}
\thanks{Denotes equal contribution}
\affiliation{Beijing National Laboratory for Condensed Matter Physics, and Institute of Physics, Chinese Academy of Sciences, Beijing 100190, China}
\affiliation{University of Chinese Academy of Sciences, Beijing 100049, China}
\author{Peipei Wang}
\affiliation{Department of Physics, Southern University of Science and Technology, Shenzhen 518055, China}
\author{Liyuan Zhang}
\affiliation{Department of Physics, Southern University of Science and Technology, Shenzhen 518055, China}
\author{Jennifer Cano}
\thanks{Corresponding authors: xu.du@stonybrook.edu, daix@ust.hk, jennifer.cano@stonybrook.edu}
\affiliation{Department of Physics and Astronomy, Stony Brook University, Stony Brook NY 11794, USA}
\affiliation{Center for Computational Quantum Physics, Flatiron Institute, New York, New York 10010, USA}
\author{Xi Dai}
\thanks{Corresponding authors: xu.du@stonybrook.edu, daix@ust.hk, jennifer.cano@stonybrook.edu}
\affiliation{Materials Department, University of California, Santa Barbara, CA 93106-5050, USA}
\affiliation{Department of Physics, The Hongkong University of Science and Technology, Clear Water Bay, Kowloon 999077, Hong Kong, China}
\author{Xu Du}
\thanks{Corresponding authors: xu.du@stonybrook.edu, daix@ust.hk, jennifer.cano@stonybrook.edu}
\affiliation{Department of Physics and Astronomy, Stony Brook University, Stony Brook NY 11794, USA}

\maketitle

\section*{Abstract}
\par
\textbf{The geometric phase (Berry phase) of an electronic wave function is the fundamental basis of the topological properties in solids. Modulating band structure provides a tuning knob for the Berry phase, and in the extreme case drives a topological phase transition. Despite the significant developments in topological materials study,  it remains a challenge to tune between different topological phases while tracing the impact of the Berry phase on quantum charge transport, in the same material. Here we report both in a magnetotransport study of ZrTe$_5$. By tuning the band structure with uniaxial strain, we directly map a weak- to strong- topological phase transition through a gapless Dirac semimetal phase via quantum oscillations. Moreover, we demonstrate the impact of the strain-tunable spin-dependent Berry phase on the Zeeman effect through the amplitude of the quantum oscillations. We show that such a spin-dependent Berry phase, largely neglected in solid-state systems, is critical in modeling quantum oscillations in Dirac bands in topological materials.}

\section*{Main}

\begin{figure*}
    \centering
    \includegraphics[width=\textwidth]{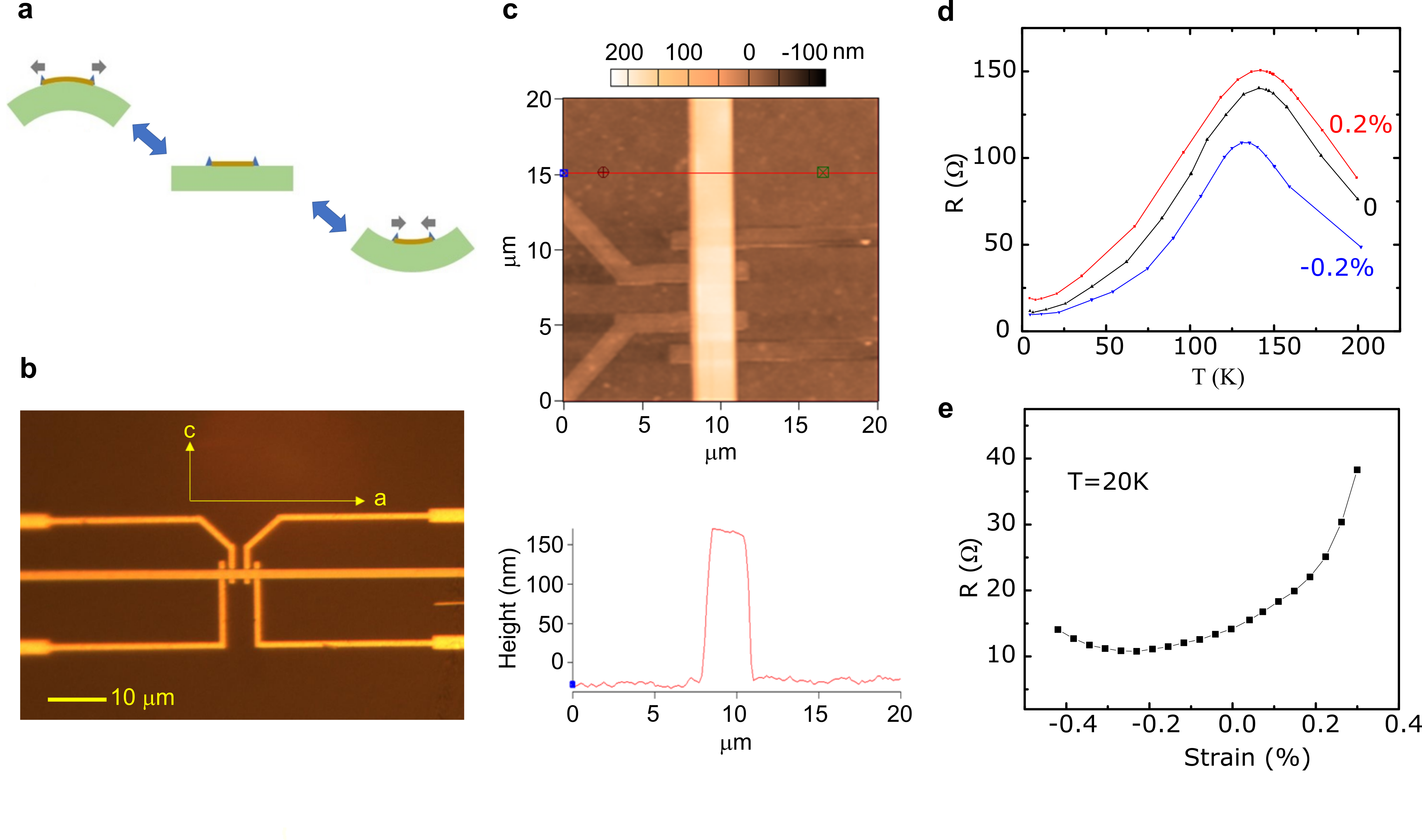}
    \caption{ZrTe$_5$ under strain. a. Application of uniaxial strain through substrate bending. b. Optical microscope image of ZrTe$_5$ microcrystal on Polyimide substrate with predefined electrodes. c. Atomic force microscope (AFM) image of the sample. The bottom panel show the cross section of the ZrTe$_5$ crystal, with a thickness of approximately 190 nm. d. Resistance versus temperature dependence under various external strains. e. Non-monotonic strain dependence of resistance at the fixed temperature of 20K, showing a resistance minimum at compressive strain $\approx -0.2\%$.}
    \label{Fig1}
\end{figure*}

Tuning the band structure of a topological material causes a continuous evolution of Berry phase along the Fermi surface and can ultimately drive a topological phase transition by closing and reopening the bulk band gap\cite{KaneMele2005,KaneMele2005-2,BHZ2006,konig2007,Moore2007,Xu2012}. Both the evolution of the band gap and the Berry phase can be probed in quantizing magnetic fields, where Shubnikov de Haas (SdH) oscillations from Landau level (LL) quantization provides a tool to probe “Fermiology”\cite{shoenberg1984,Mikitik1999,Novoselov2005,Zhang2005}. 
In addition to LL quantization, the Zeeman effect splits the spin-degenerate LLs and reduces the SU(2) Berry phase to the U(1) Berry phases, which play a crucial role in the SdH oscillations.
Interestingly, while the Zeeman effect has been extensively studied over the past decades\cite{luttingerperturbation,cohengfactor1960}, its connection to band topology and non-trivial Berry phase has emerged only recently. A Dirac band with a finite mass gap hosts spin-dependent Berry phase which can modify the SdH oscillations through the Zeeman effect, as recently demonstrated in the spin-zero effect\cite{Wang9145, PhysRevB.101.125118}.  Such a spin-dependent Berry phase is expected to be fundamentally generic for Dirac-like bands, and is band parameter dependent. It is therefore desirable to conduct a comprehensive study of such effect in a system with tunable band parameters.

\begin{figure*}
    \centering
    \includegraphics[width=\textwidth]{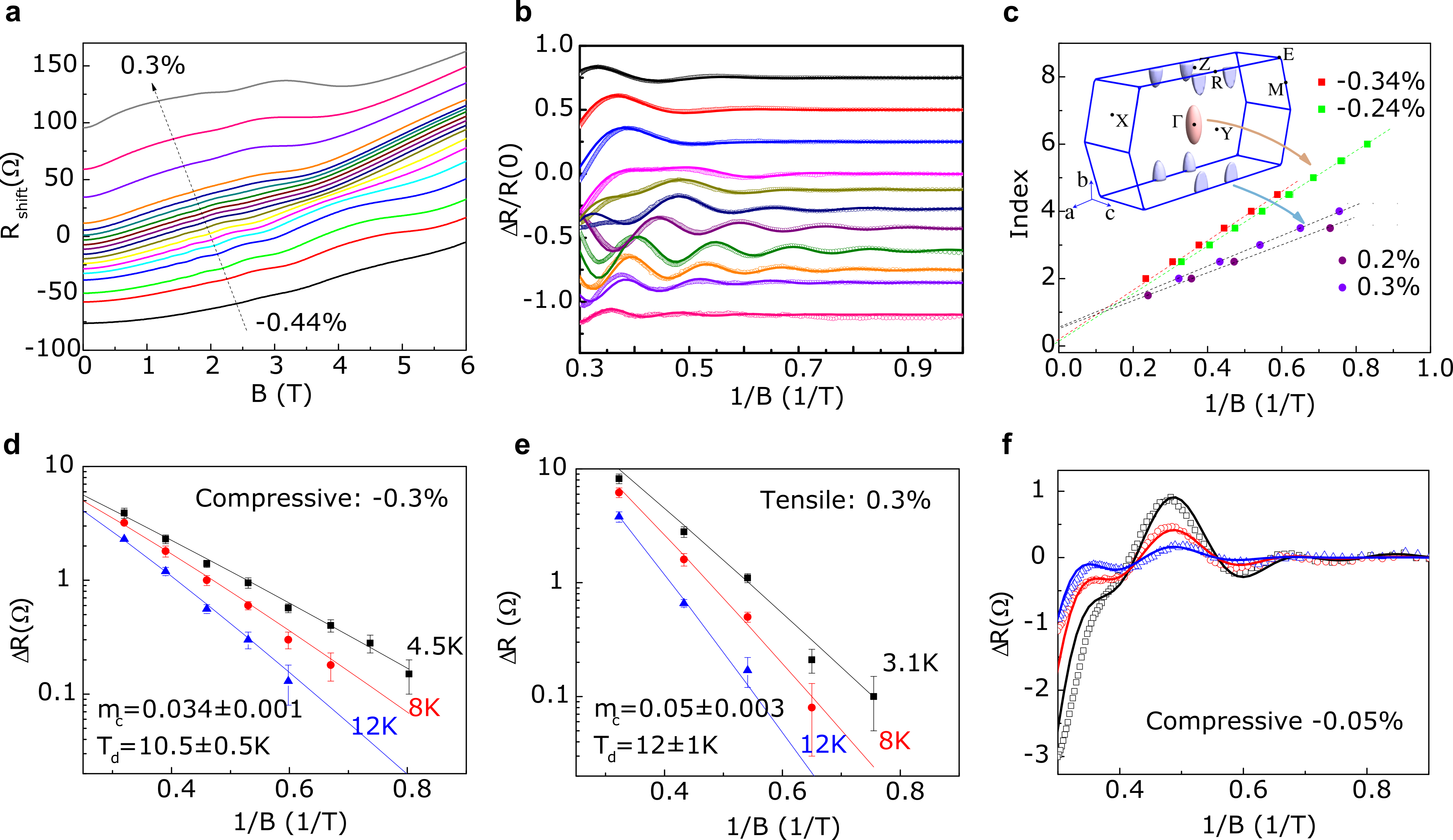}
    \caption{Magnetotransport measurements on ZrTe$_5$ under strain. a. Resistance versus magnetic field in various external strains. The curves are shifted in proportion to the strain for clarity, from the bottom to the top: -0.44, -0.34, -0.3, -0.24, -0.21, -0.19, -0.17, -0.15, -0.13, -0.11, -0.09, -0.07, -0.05, -0.03, 0, 0.1, 0.2, 0.3\%. b. SdH oscillations versus inverse magnetic field in various external strains. The curves ares shifted in proportion to the strain for clarity, from the bottom to the top: -0.44, -0.34, -0.3, -0.24, -0.17, -0.11, -0.05, 0, 0.1, 0.2, 0.3\%. Symbols are experimental data and solid lines are simulations using the two-component LK formula. c. Landau index, obtained from SdH oscillations, versus inverse magnetic field in large compressive and tensile strains. Here resistance peaks and valleys are assigned integer and half integer indices, respectively. The inset show the pockets of Fermi surface, with the Dirac band at the $\Gamma$ point and a parabolic side band between the R and E points dominating the SdH oscillations under strong compressive and tensile strains, respectively. d, e. SdH amplitude versus inverse magnetic field and temperature in strong compressive (-0.3\%) and tensile (0.3\%)strains. The symbols are experimental data and the solid lines are fittings using the LK formula. From the fittings, the effective cyclotron mass and Dingle temperature can be obtained. f. Under medium strains the SdH oscillations (open symbols) are directly simulated (solid lines) with the two-component LK formula at various temperatures: black: 3K; red: 8K; and blue: 12K.}
    \label{Fig2}
\end{figure*}

Despite the significant developments in topological materials study in recent years, it remains a challenge to tune between different topological phases, and correspondingly the Berry phase, in the same material. The transition metal pentatelluride ZrTe$_5$ is a promising material for such study. ZrTe$_5$ is a van der Waals layered material with layer planes extending along the a- and c- lattice directions and stacking along the b direction. It hosts intriguing properties such as a resistance peak (Lifshitz transition) \cite{NatComm.15512, Chi2017}, chiral magnetic effect \cite{Li2016}, and 3D quantum Hall effect\cite{Tang2019}. In its 3D bulk form, ZrTe$_5$ has a Dirac-like low energy band structure, with sample-dependent mass gap rendering the material from Dirac semimetal to topological insulator (TI) \cite{Liu2016, Li2016, Chen816,PhysRevB.96.041101, PhysRevLett.121.187401,PhysRevB.95.195119},  suggesting extreme sensitivity to lattice deformations. Recently a strain-induced weak TI (WTI) to strong TI (STI) transition has been proposed \cite{PhysRevX.4.011002,SciRep.7.45667}, followed by experimental evidence in angle-resolved photoemission spectroscopy study \cite{Zhang2021}, and indirect charge transport evidences via the chiral anomaly effect \cite{Mutcheaav9771}. A charge transport study of the quantum oscillations which directly map the topological phase transition and reveal the spin-dependent Berry phase over tunable band parameters, however, is still lacking.  

In this work, we study charge transport and SdH oscillations in ZrTe$_5$ under tunable uniaxial strain. In a magnetic field perpendicular to the a-c plane and the applied current, SdH oscillations and their evolution over strain allow direct mapping of the WTI-STI transition through the closing and reopening of the Dirac band mass gap. The dependence of the SdH oscillation amplitude on the Fermi energy and Dirac mass gap, tunable through the strain, is analyzed and compared with the quantum oscillation theory. Our results reveal that the spin-dependent Berry phase intrinsic to the Dirac bands, which has been largely neglected in previous studies, is critical in modeling the SdH oscillations in such topological materials.

The samples studied in this work are ZrTe$_5$ microcrystals mechanically exfoliated onto flexible Polyimide substrates (Fig. 1a-c), which allow application of external tensile and compressive strains along the a-axis through substrate bending, over a wide temperature range from room temperature down to below 4K (see Methods).  Our ZrTe$_5$ exhibits a strain-tunable resistance peak at $T_p\approx 140K$(Fig.1d). At $T=20K$ a non-monotonic resistance versus strain dependence is observed (Fig.1e),  consistent with the report in macroscopic ZrTe$_5$ crystals\cite{Mutcheaav9771}. A large strain gauge factor ranging $10^2-10^3$ generally presents throughout the temperature range up to room temperature, suggesting strain-tuning of the low energy electronic spectrum in ZrTe$_5$.

Next, we characterize the electrical resistance in magnetic field perpendicular to the a-c crystal plane. At the base temperature of $\approx4$K, SdH oscillations are clearly visible on the magnetoresistance background and evolve with changing strain (Fig.2a). Plotting the oscillatory part $\Delta R$ versus inverse magnetic field $1/B$, equally spaced resistance oscillations can be resolved (Fig.2b). In the limits of large compressive and tensile strains applied here, the oscillation amplitude monotonically decreases with increasing 1/B. Under mild compressive strains, however, the SdH oscillation amplitude appears non-monotonic, suggesting contributions from more than a single Fermi surface. 

\begin{figure*}
    \centering
    \includegraphics[width=\textwidth]{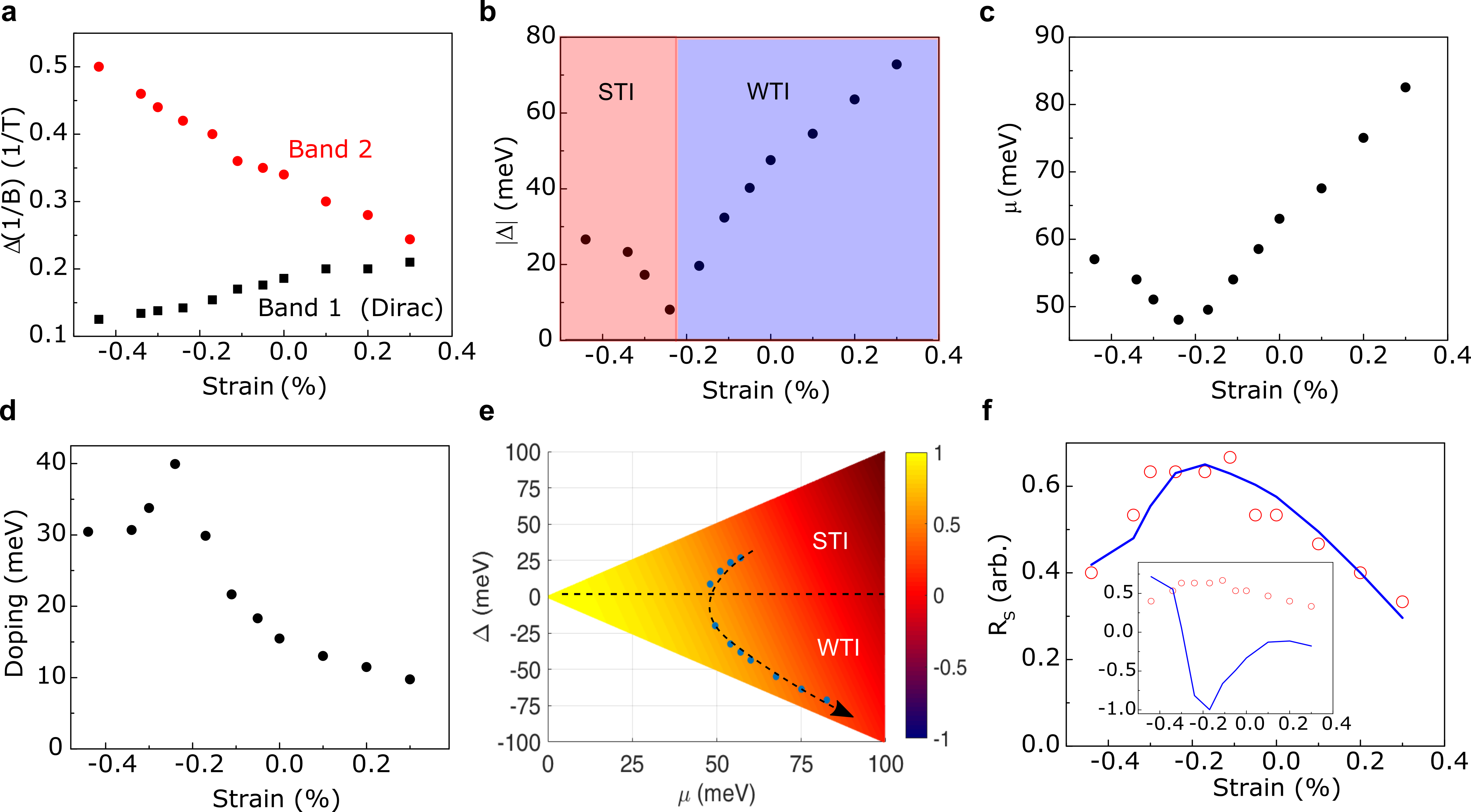}
    \caption{Strain tuning of band structure and spin-dependent Berry phase. a. Strain dependence of SdH oscillation period in inverse magnetic field, for the Dirac band (band 1) and the trivial band (band 2). The results are obtained by simulating the SdH oscillations. b, c. Strain tuning of mass gap size and Fermi energy. The mass gap and the Fermi energy are calculated from Eqs. (2) and (3), with $m_c$ and $E_F$ obtained from 2-component LK formula simulations of the the magnetic field and temperature dependence of the SdH oscillation curves. Tuning the external strain from compressive to tensile, the mass gap closes and reopens at around $-0.2\%$, consistent with the STI-WTI topological phase transition. d. Strain dependence of doping energy, the energy of the Fermi surface counted from the bottom of the conduction band. e. Theoretical intensity plot of SdH oscillation amplitude reduction factor from Zeeman effect, $R_S$, as functions of Fermi energy and mass gap. The dotted line shows the boundary between the STI phase (positive mass gap) and the WTI phase (negative mass gap). The symbols in the mass gap-Fermi energy parameter space correspond to the experimental data points at various external strains, evolving following the arrow from compressive to tensile. By convention we assign a negative sign to the mass gap on the WTI side with external strain larger than $\approx -0.2\%$. f. Theory(solid lines)/experiment(open symbols) comparison of $R_S$ as functions of mass gap and Fermi energy at various strains. The experimental values of $R_S$ are scaled by a common constant factor of 0.6 for comparison with the theory. The theoretical model qualitatively reproduces the experiment results by considering the spin-dependent Berry phase. By contrast, the inset shows the theoretical curve without considering the spin-dependent phase, which fails to explain the experimental data.}
    \label{Fig3}
\end{figure*}

Our analysis of the SdH oscillations focuses on mapping out the strain-dependent mass gap, Fermi level, and the spin-dependent Berry phase. We model ZrTe$_5$ with the anisotropic Dirac Hamiltonian (see Methods). In quantizing magnetic fields, the extremal cross-sectional area of the Fermi surface is $S_F = \pi (\mu^2 - \Delta^2) / (\hbar^2 v^2)$. The corresponding cyclotron mass is $m_c = \frac{\hbar^2 \partial S_k}{2 \pi \partial
\mu} = \frac{\mu}{v^2}$. Here $\mu$ is the Fermi energy, $\Delta$ is the mass gap, and $v$ ($\approx 5\times 10^5m/s$\cite{Liu2016}) is the Dirac band velocity in the a-c plane. The SdH oscillations can be modeled with the Lifshitz–Kosevich (LK) formula: to the first order, each single band contributes to the SdH oscillations: 
\begin{equation}
\Delta R/R_0 \propto R_D R_T  \cos \left( \frac{\hbar S_{F }}{e B} + \phi_{B }
   + \pi + \delta\right)
\label{tdepeq}
\end{equation}
We note that the conventional (parabolic band) LK formula and its Dirac version \cite {PhysRevB.71.125124} have the same mathematic form with their corresponding band parameters (see Supplementary Information). In Eq.(1) $R_0$ is the zero magnetic field resistance. Defined for Dirac band, $R_D=\exp{(-2\pi^2k_B T_d |\mu|/\hbar e B v^2)}$ is the amplitude reduction factor from disorder scattering, where $T_d$ is the Dingle temperature characterizing the level of disorder,and $k_B$ is the Boltzmann constant. $R_T=\xi/\sinh{(\xi)}$ is the amplitude reduction factor from temperature, where $\xi=2\pi^2 k_B T \mu/\hbar e B v^2$.  $\phi_B$ is the Berry phase, and $\delta=\pm\pi/4$ in 3D materials.

Now we consider the Zeeman effect, which splits a spin-degenerate band into two, each with a different Fermi surface extremal cross-sectional area: $S_{F \uparrow / \downarrow} = S_F \pm \alpha B$, and Berry phase: \ $\phi_{B \uparrow
/ \downarrow} = \phi_B \pm \phi_s$ \cite{PhysRevB.101.125118}. Here $\alpha$ describes the splitting of extremal cross-sectional area of the Fermi surface and $\phi_s$ is the spin-dependent part of the Berry phase. The overall quantum oscillation is a summation of the two oscillation terms from the two spin bands: 
$\cos \left( \frac{\hbar S_{F \uparrow}}{e B} + \phi_{B \uparrow}
   + \pi +\delta \right) + \cos \left( \frac{\hbar S_{F \downarrow}}{e B} + \phi_{B
   \downarrow} + \pi + \delta \right) \\= 2\cos \left( \frac{\hbar S_F}{e B} + \phi_B + \pi + \delta \right) \cos \left( \frac{\hbar \alpha}{e} +\phi_s \right)$
Interestingly $S_F$ and $\phi_B$, which are the common (spin-independent) part of the two spin bands before the
Zeeman splitting, determines the period and phase offset of the quantum oscillations; while the difference between the two spin bands determines the amplitude. 

For systems with nontrivial Berry curvature, the Berry phase splitting $\phi_s$ along the Fermi surface will be nonzero. In particular, for a Dirac electron system, we find (see Methods)
\begin{eqnarray}
& \phi_s = \pi\frac {\Delta}{\mu} \\
  & \phi_{B \uparrow} = \pi + \phi_s = \pi \left( 1 +
  \frac{\Delta}{\mu} \right) &  \nonumber\\
  & \phi_{B \downarrow} = \pi - \phi_s = \pi \left( 1 -
  \frac{\Delta}{\mu} \right) & 
\end{eqnarray}
Defining the Zeeman effect-induced SdH amplitude reduction factor: $R_s=\cos \left( \frac{\hbar \alpha}{e } +  \phi_s \right)$, we derive for the Dirac band: 

\begin{equation}
  \Delta R/R_0 =A R_D R_T R_s \cos \left( \frac{\pi (\mu^2 - \Delta^2)}{e B \hbar^{} v^2} +\delta \right) ,
  \end{equation}
Further, the coefficient $\alpha$ is computed from the g-factor tensor obtained from first principle calculations at given values of $\Delta$ and $\mu_F$.

 \begin{figure*}[tp]
    \centering
    \includegraphics[width=\textwidth]{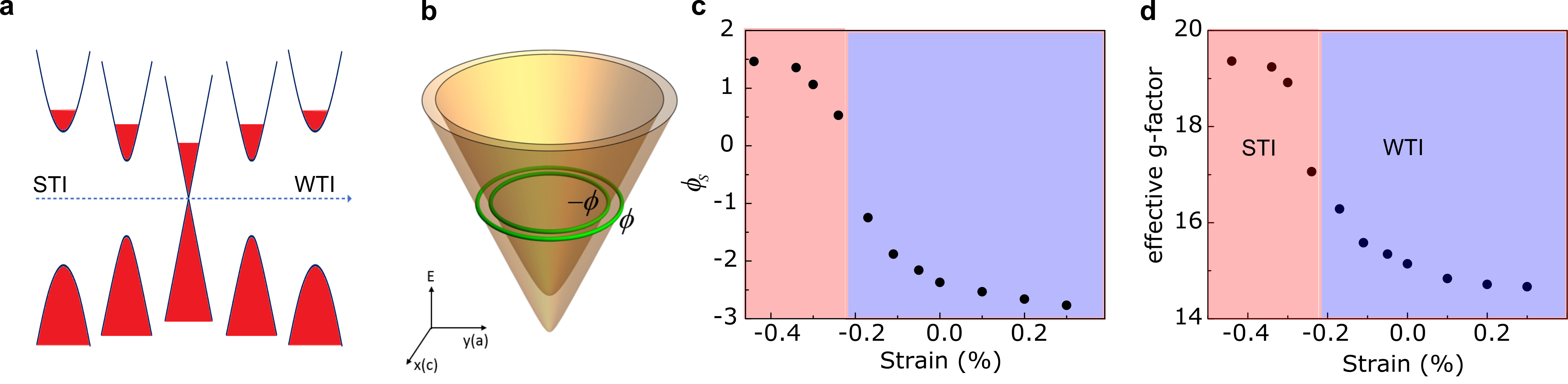}
    \caption {Evolution of band structure and Zeeman effect over strain. a. Evolution of Dirac band and Fermi level through the STI to WTI phase transition with increasingly tensile strain. b. Schematic energy dispersion with $k_z=0$ and magnetic field in the z(b) direction. Green circles indicate the boundary of extremal cross-sections for the spin bands.  The spin-dependent Berry phases are $\phi$ and $-\phi$ over the two circles. c. spin-dependent Berry phase $\phi_s$ as a function of strain, calculated with mass gap and Fermi energy obtained from the SdH oscillations. d. Strain tuning of effective g-factor.}
    \label{Fig4}
\end{figure*}

Under large compressive and large tensile strains, we observed approximately single SdH oscillation frequencies and a monotonic decay of SdH oscillation amplitude over increasing $1/B$, suggesting that SdH oscillations are dominated by single bands there.  The linear dependence of the LL indices on $1/B$ (Fig. 2c) extrapolates to the index values of close to 0 and 1/2 at $1/B = 0$,  corresponding to a SdH oscillation phase change from approximately zero to $\pi$. Fitting the $1/B$ and $T$ dependence of the SdH oscillation amplitude to the LK formula (examples shown in Fig.2d-f) provides estimations to the cyclotron mass and the Dingle temperature at these particular strains. 

More generally, the SdH oscillations show a complex $1/B$ dependence which can be modeled considering contributions from two bands: $\Delta R=\Delta R_1+\Delta R_2$. Here band 1 is a Dirac band with a common Berry phase of $\pi$ for its two spins bands and hence a overall SdH phase which is much smaller than $\pi$ (Fig.~2c). Band 2, a ``trivial'' parabolic band possess a overall SdH phase which is close to  $\pi$. We note that evidence of charge transport contribution from a secondary band has been reported previously\cite{Chi2017}. The change of SdH phase shown in Fig.~2c is due to the strain-evolution of the relative contribution from the two bands, instead of band topology transition within the same band. 

Fig. 2b compares the measured SdH oscillations with the two-component LK formula simulations at base temperatures. Here the Dirac band SdH oscillations are modeled with Eq.(4), while the SdH oscillations from the trivial band are modeled with the conventional LK formula, with parameters including $T_d$, Fermi energy and cyclotron mass $m_c$. The simulation reproduces the experimental observations, with small deviations which may be attributed to the choices of background curves when extracting the oscillatory part of the data. The more general comparisons including temperature dependence are shown in the Supplementary Information. Fig.~3a-c  plots the key simulation parameters. Focusing on the the Dirac band, a closing and re-opening of mass gap with tuning external strain happens at a compressive strain of $\approx -0.2\%$. This, remarkably, is a direct transport evidence of the theoretically predicted WTI to STI transition. Associated with such transition, the Fermi energy (measured from the center of the mass gap) shows a non-monotonic strain dependence, reaching a minimum close to the transition strain where mass gap vanishes. From the mass gap and the Fermi energy, we calculate the electron doping: $E_F-\Delta$, which characterizes the energy of the Fermi level in relation to the bottom of the conduction band. The result indicates a maximum electron doping at the WTI-STI transition, which decreases when strain-tuned away from the transition. Generally the analysis of the band parameters suggests a band evolution with strain which is depicted in Fig.~4a.

Next we focus on the impact of Zeeman effect on SdH oscillation amplitude ($R_s$). Experimentally we obtain $R_s$ at every strain by quantitative simulation of the SdH oscillations. We then compare the results with the theoretical expectation: $R_s=\cos \left( \frac{\hbar \alpha}{e } +  \phi_s \right)$. Here $\frac{\hbar\alpha}{e}$, which is associated with the splitting of extremal cross-sectional area of the Fermi surface, follows (see Methods):
\begin{equation}
    \frac{\hbar\alpha}{e}
    = - \pi \frac{\Delta}{\mu} - \pi\frac{\hbar^2}{m_e}\frac{\mu}{v^2}\left(\frac{g_{p,z}}{2}(1+\frac{\Delta}{\mu})-\frac{g_{s,z}}{2}(1-\frac{\Delta}{\mu})\right) \label{eq:halpha}
\end{equation}
We adopt the g-factors for the $s$ and $p$ orbitals $g_{p,z} = 9.66, g_{s,z} = -6.45$, as computed from first-principle calculations (DFT-mBJ) \cite{song2021first}. The spin dependent part of the Berry's phase $\phi_s=\pi \frac{\Delta}{\mu}$ is calculated from the mass gap and Fermi energy obtained from simulating the SdH oscillations. The theory shows good qualitative agreement with the data on the strain dependence of $R_s$. By contrast, the conventional modeling of Zeeman effect on SdH oscillations which only considers the Fermi surface splitting (Eq.~(5)) completely fails to match with the experimental observations. This comparison definitively highlights the importance of spin-dependent Berry phase in the Zeeman effect.

The significant strain-tunability of the spin-dependent Berry phase $\phi_s$ is shown in Fig.4c. Accompanied by the topological phase transition where the mass gap vanishes, $\phi_s$ passes through zero when the system is in the Dirac semimetal phase. We also compute the effective g-factor by comparing $R_s$ with the conventional Zeeman effect induced SdH amplitude reduction factor $\cos{(\pi g m_c/(2m_e))}$\cite{shoenberg1984,Liu2016}. This leads to $g=2m_e(\hbar \alpha/e + \phi_s)/(\pi m_c)$, whose strong strain tunability is shown Fig.4d.

In conclusion, we have carried out a magnetotransport study on the evolution of band topology and non-trvial Berry phase over strain-tunable band parameters in ZrTe$_5$. The strain-dependent SdH oscillations allow direct mapping of the closing and reopening of the Dirac band mass gap, which is consistent with a WTI-STI transition in ZrTe$_5$. Moreover we observed the non-trivial Berry phase and its dependence on band parameters over the transition. Such a spin-dependent Berry phase is generic and intrinsic to Dirac band structure, and is a critical factor in modeling Zeeman effect in SdH oscillations in topological materials.

\section*{Methods}
\subsection*{Sample fabrication and characterization}
The samples studied in this work are ZrTe$_5$ microcrystals mechanically exfoliated onto flexible Polyimide substrates (Fig. 1a,b), which allow application of tensile and compressive strains over a wide temperature range from room temperature down to ~4K. ZrTe$_5$ single crystals were synthesized by chemical vapor transport (CVT) method, with iodine as transport agency.
Stoichiometry amounts of Zr(4N) and Te(5N) powder, together with 5mg/mL I$_2$, were
loaded into a quartz tube under argon atmosphere. The quartz tube was flame sealed
and then placed in a two-zone furnace, a temperature gradient from 480°C to 400°C was
applied. After 4 weeks reaction, golden, ribbon-shaped single crystals were obtained, of typical size about $0.6\times0.6\times5mm$. The X-ray diffraction characterization of the crystals is shown in the Supplementary Information. 

To avoid degradation of the material from ambient exposure, the crystals are exfoliated and press-contacted on predefined gold electrodes on 120$\mu$m-thick Polyimide substrates in Ar environment, and are encapsulated with poly(methyl methacrylate) (PMMA) which both protects the crystals from  degradation and facilitates the application of strain. A detailed description on the sample fabrication and on the contact/crystal interface can be found in reference\cite{Mills2019}. All measurements were done with electric current applied along the a-axis. Typical contact resistance over a ~10 $\mu m^2$ contact area is between few hundred ohms to a few kilo-ohms, and does not change significantly over the application of strain through substrate bending. Uniaxial strain is applied to the samples using a 4-point bending setup, with motorized precision control to the substrate curvature\cite{Guan107.193102,Guan2017}. Because the substrate is much thicker than the microcrystals, significant uniaxial strain can be achieved at relatively mild substrate curvatures under which the strain on the microcrystals is predominantly  tensile/compressive due to the elongation/compression of the top substrate surface where the microcrystals are PMMA-pinned. Besides external strain applied through substrate bending, both the crystals, the substrates, and the encapsulating PMMA layer go through thermal expansion/contraction upon temperature change. Because of such complication it is difficult to precisely characterize the absolute strain on the crystals. Here we focus on the dependence of the transport characteristics on the change of strain induced by substrate bending, which is described in the discussion below as “external strain”. In magnetotransport measurements, we limit the temperature range between 3-20K, where the change of thermal expansion is small in comparison to the range of the external strain. All charge transport measurements were performed in a Oxford Variable Temperature Insert (VTI) with a superconducting magnet, using standard lock-in technique.

\subsection*{Computing the \texorpdfstring{$g$}{lg}-factor}

To compute the effect of Zeeman coupling on quantum oscillations, we study the $\mathbf{k}\cdot \mathbf{p}$ Hamiltonian near $\Gamma$, which is an anisotropic gapped Dirac Hamiltonian \cite{song2021first}:
\begin{equation}
H(\mathbf{k} )  = 
\begin{pmatrix}
\Delta \sigma_0 & \sum_i v_i k_i \sigma_i \\
\sum_i v_i k_i \sigma_i & -\Delta \sigma_0
\end{pmatrix},
\label{eq:H0}
\end{equation}
in the basis $|p,\frac{1}{2} \rangle$, $|p, -\frac{1}{2}\rangle$, $|s, \frac{1}{2}\rangle$, $|s, -\frac{1}{2} \rangle$, where $\pm \frac{1}{2}$ refers to the $z$-component of spin.
In Eq.~(\ref{eq:H0}), $\Delta$ is half the mass gap; $v_{x,y,z}$ are the anisotropic Dirac cone velocities; $\sigma_{x,y,z}$ are the Pauli matrices; and $\sigma_0$ is the $2\times 2$ identity matrix.
The eigenvalues of $H(\mathbf{k})$ are 
\begin{equation}
E_\pm = \pm \sqrt{ \Delta^2 + \sum_i v_i^2 k_i^2},
\label{eq:E0}
\end{equation}
where all bands are doubly-degenerate due to the combination of time-reversal and inversion symmetry.
The eigenvectors are given by
\begin{equation}
| \psi^\pm_{m,\mathbf{k}} \rangle = \sqrt{ \frac{ E_\pm + \Delta}{2E_\pm } } \begin{pmatrix}
u_m \\ \frac{v}{E_\pm + \Delta} \bm{p \cdot \sigma} u_m \end{pmatrix},
\label{eq:eigenstates}
\end{equation}
where $u_1= (1,0)^T$, $u_2 = (0,1)^T$
and we have defined the rescaled coordinates $p_i = v_i k_i /v$.

The Zeeman coupling for the four-band model is:
\begin{equation}
H^Z_0 = 
\mu_B \begin{pmatrix}
\sum_i g_i^p B_i & 0 \\ 
0 & \sum_i g_i^s B_i
\end{pmatrix},
\label{eq:HZ0}
\end{equation}
where $g_{x,y,z}^{s,p}$ are the matrix-valued $g$-factors for the $s$ and $p$ orbitals, which are anisotropic due to the crystal symmetry and which differ from their bare value due to coupling with all the other bands not in the four-band model.

To compute the quantum oscillations, we must downfold this four-band model into the two bands at the Fermi level.
We seek an \textit{effective} $g$-factor for the two degenerate bands that make up the Fermi surface, such that when projected onto those bands, the Zeeman Hamiltonian takes the form: 
\begin{equation}
H^Z_{mm'} = \mu_B \mathbf{g}_{mm'} \cdot \mathbf{B}, 
\end{equation}
where $\mu_B = \frac{e\hbar}{2m_e}$ is the Bohr magneton and $m,m'$ index the two bands at the Fermi surface. 

The effective $g$-factor has two contributions: 
\begin{align}
    \bm{g} = \bm{g}_0 + \bm{g}_D,
    \label{eq:gdecompose}
\end{align}
where $\bm{g}_0$ is the projection of $g^s$ and $g^p$ (defined by $H^Z_0$) onto the two conduction bands at the Fermi level and $\bm{g}_D$ is an extra orbital contribution that we will describe below.
For a magnetic field in the $z$ direction, using the expressions for the conduction band eigenstates from Eq.~(\ref{eq:eigenstates}),
\begin{align}
\bm{g}_{0,z} 
 &= \left( \frac{E_+ + \Delta}{2E_+} \right)\left( g_z^p  +  g_z^s \frac{v^2(p_z^2 - p_x^2 - p_y^2)}{(E_+ + \Delta)^2}  \right) \sigma_z,
\end{align}
where the Pauli matrix $\sigma_z$ acts in the basis of the two bands at the Fermi level.
Ultimately, we will only need the $g$-factor at the extremum of the Fermi surface, where $p_z=0$ and $v^2(p_x^2 + p_y^2) = \mu^2 - \Delta^2$. In this case,
\begin{equation}
\bm{g}_{0,z}^\text{ext} = \frac{1}{2}\left(g_z^p \left(1+\frac{1}{\gamma}\right)-g_z^s \left(1-\frac{1}{\gamma}\right)\right) \sigma_z
\label{eq:gelse}
\end{equation}
where $\gamma = \frac{\mu}{\Delta}$ is a dimensionless constant.

Returning to the second term in Eq.~(\ref{eq:gdecompose}),
$\bm{g}_D$ is the orbital contribution to the $g$-factor from the Dirac cone~\cite{Xiao2010Berry}:
\begin{equation}
\mathbf{g}_{D,mm'}(\mathbf{k})  = \frac{im_e}{\hbar^2}\sum_{ijk}\varepsilon_{ijk}\hat{\mathbf{e}}_k \langle \partial_i \psi_{m,\bm{k}}| \left( H(\bm{k}) - E_{m,\bm{k}} \right) |\partial_j \psi_{m',\bm{k}}\rangle ,
\label{eq:gfactor2}
\end{equation}
where $\psi_{m,\mathbf{k}} $ are the eigenvectors of Eq.~(\ref{eq:H0}).

The first step to evaluate $\bm{g}_D$ is to find the derivatives of the eigenstates:
\begin{align}
    \frac{\partial |\psi_{m,\bm{k}}\rangle}{\partial p_i} 
    &= \sqrt{\frac{E+\Delta}{2E}}\left( \begin{array}{c}
        -\frac{\Delta v^2 p_i}{2E^2(E+\Delta)}u_m \\ 
        -\frac{(\Delta+2E)v^3 p_i}{(E+\Delta)^22E^2}\bm{p \cdot \sigma}u_m+\frac{v}{E+\Delta}\sigma_iu_m
    \end{array}\right),
    \label{eq:deigenstates}
\end{align}
where we have dropped the subscript/superscript $\pm$ on $E$ and $\psi$ to reduce clutter.
After some algebra, it follows that:
\begin{widetext}
\begin{align}
    \left(H-E_{m,\bm{k}}\right)\frac{\partial |\psi_{m,\bm{k}}\rangle}{\partial p_i} 
    &=\sqrt{\frac{E+\Delta}{2E}} \frac{1}{E(E+\Delta)}\left(\begin{array}{c}
        \Delta v^2p_i u_m + iEv^2\sum_{jk}\varepsilon_{jik}p_j\sigma_ku_m\\ 
        v^3p_i \bm{p\cdot\sigma}u_m - E(E+\Delta)v\sigma_iu_m
    \end{array}\right),
\end{align}
which yields the matrix element
\begin{align}
    \langle \frac{\partial \psi_{m,\bm{k}}}{\partial p_i}|H(\bm{k}) - E_{m,\bm{k}}|\frac{\partial \psi_{m',\bm{k}}}{\partial p_j}\rangle 
    &=  \frac{v^2}{2E^2} u_m^\dagger \left[  \frac{-i v^2\varepsilon_{jkl}p_ip_k  \sigma_l  + iv^2 \varepsilon_{ikl} p_jp_k \sigma_l}{E+\Delta}  - i E \varepsilon_{ijk} \sigma_k  + \frac{ v^2 p_i p_j }{E} - E \delta_{ij}\right]u_{m'}
\end{align}
\end{widetext}
The terms symmetric under the exchange of the $i$ and $j$ indices (i.e., the last two terms) will cancel in the sum in Eq.~(\ref{eq:gfactor2}). Thus, applying Eq.~(\ref{eq:gfactor2}), the extra orbital contribution to the $g$-factor is given by:
\begin{widetext}
\begin{align}
    \bm{g}_{D} 
    &=\frac{g^L}{2\tilde{\gamma}^2}\sum_{ijk} \left[\tilde{\gamma} \varepsilon_{ijk}\varepsilon_{ijk}\sigma_k-\frac{2v^2/\Delta^2}{1+\tilde{\gamma}}\sum_{qr}\varepsilon_{iqr}\varepsilon_{ijk}p_qp_j\sigma_r\right]\frac{v_iv_j}{v^2}\hat{\bm{e}}_k 
    \label{eq:gfactorresult}
\end{align}
\end{widetext}
where  $g^{L} \equiv \frac{m_ev^2}{\hbar^2\Delta}$ and $\tilde{\gamma} \equiv \frac{E}{\Delta}$ are dimensionless constants.

We now simplify this result by taking $B$ in the $z$-direction and restricting to the boundary of the extremal cross-section, where $p_z = 0$ and $v^2(p_x^2+p_y^2)=\mu^2-\Delta^2$. Plugging this into Eq.~(\ref{eq:gfactorresult}), we obtain the $z$-component of $g$:
\begin{equation}
\bm{g}_{D,z}^\text{ext} = \frac{g^L}{\tilde{\gamma}^2}\frac{v_xv_y}{v^2}\sigma_z
\label{eq:gdirac}
\end{equation}

The total effective $g$-factor at the extremal cross-section of the Fermi surface is found by combining Eqs.~(\ref{eq:gelse}) and (\ref{eq:gdirac}):
\begin{equation}
    \bm{g}_{z}^\text{ext} = \left[\frac{m_e\Delta v_xv_y}{\hbar^2\mu^2}+\frac{g_z^p}{2}\left(1+\frac{\Delta}{\mu}\right)-\frac{g_z^s}{2}\left(1-\frac{\Delta}{\mu}\right) \right]\sigma_z 
    \label{eq:gextfinal}
\end{equation}
This expression shows that the \textit{g}-factor in the $\hat{z}$-direction is independent of momentum on the boundary of the extremal cross-section and is diagonal, i.e., $|\psi_{m,\bm{k}}\rangle$ are eigenstates of the Zeeman coupling.

\subsection*{Quantum oscillations}
We now discuss the quantum oscillations of the anisotropic Dirac semi-metal. 
As discussed in the main text, the quantum oscillations for a doubly degenerate electronlike Fermi surface are proportional to
\begin{align}
    \Delta R/R_0 &\propto \cos \left( \frac{\hbar S_F}{e B} + \phi_B + \pi + \delta \right) \cos \left( \frac{\hbar \alpha}{e} +\phi_s \right)
\label{eq:LK}
\end{align}
where $\phi_{B,\uparrow / \downarrow} = \phi_B \pm \phi_s$ are the Berry phases around the extremal Fermi surface for each of the two spins and $S_{F,\uparrow / \downarrow} = S_F \pm \alpha B$ are the extremal areas of the Fermi surface for each spin.

The Berry connection is defined by \cite{Xiao2010Berry}: $\bm{A}_{mm'}(\bm{k}) = \sum_k i\langle \psi_{m,\bm{k}}|\frac{\partial}{\partial k_k}|\psi_{m',\bm{k}}\rangle \bm{e}_k $.
By plugging in the expressions for the eigenstates and derivatives of eigenstates in Eqs.~(\ref{eq:eigenstates}) and (\ref{eq:deigenstates}), we find the explicit expression:
\begin{multline}
 \bm{A}(\bm{k}) =\frac{v}{\Delta^2}\frac{1}{2\tilde{\gamma}(\tilde{\gamma}+1)}\left[(p_y\sigma_z-p_z\sigma_y)v_x\bm{e}_x+ \right.\\
 \left. (p_z\sigma_x-p_x\sigma_z)v_y\bm{e}_y+(p_x\sigma_y-p_y\sigma_x)v_z\bm{e}_z \right]
 \end{multline}
To find the Berry phase, we ultimately need $\bm{A}(\bm{k})\bm{\cdot}\mathrm{d}\bm{k}$
on the extremal cross-section, where $p_z=0$ and $\mathrm{d}\bm{k}$ is in the $x-y$ plane, which yields:
\begin{align}
    \bm{A}(\bm{k})\bm{\cdot}\mathrm{d}\bm{k} 
    &=\frac{v_xv_y}{\Delta^2}\frac{\sigma_z}{2\tilde{\gamma}(\tilde{\gamma}+1)}\left(k_y\bm{e}_x-k_x\bm{e}_y\right)\bm{\cdot} \mathrm{d}\bm{k},
\end{align}
where $\sigma_z$ is acting in the space of the two conduction bands at the Fermi level.
We find the Berry phase around each extremal ring of the Fermi surface by integrating over the the boundary of the extremal cross-section:
\begin{equation}
\oint \bm{A}(\bm{k})\bm{\cdot}\mathrm{d}\bm{k}  \!=\! \frac{v_xv_y}{2 \Delta^2\tilde{\gamma}(\tilde{\gamma}+1)}\oint \left(k_y\bm{e}_x-k_x\bm{e}_y\right)\bm{\cdot} \mathrm{d}\bm{k}\label{eq:phi}
 \end{equation}
The loop integral can be turned into an area integral using Stokes theorem:
\begin{align}
 \iint \nabla_{\bm{k}}\times \left(k_y\bm{e}_x-k_x\bm{e}_y\right) \cdot \mathrm{d}\bm{S}_{\bm{k}} &= -2 \iint d S_\mathbf{k} \\&= - 2\pi \frac{\mu^2 - \Delta^2}{v_y v_x},
\end{align}
which yields:
\begin{equation}
\phi_{B,\uparrow/\downarrow} = \pi\left(1 \pm \frac{\Delta}{\mu}\right),  \label{eq:berryphase}
\end{equation}
where we have added $2\pi$ in order to ensure the Berry phase in the range of $0$ to $2\pi$.

The last term that we need to evaluate the Lifschitz-Kosevich formula is the area of the extremal Fermi surfaces.
Since the effective $g$-factor at the extremal Fermi surfaces (Eq.~(\ref{eq:gextfinal})) is diagonal in the basis of the two conduction bands, the two extremal Fermi surfaces satisfy the equation:
\begin{equation}
\mu = \sqrt{ \Delta^2 + v_x^2 k_x^2 + v_y^2 k_y^2 } \pm \mu_B \bm{g}_z^\text{ext} B,
\end{equation}
accounting for the fact that the magnetic field is in the $z$ direction and the extremal Fermi surfaces have $p_z = 0$.
Thus, each extremal cross-section forms an ellipse whose area is given by:
\begin{equation}
S = \frac{\pi}{v_x v_y} \left[ \left( \mu \mp \mu_B \bm{g}_z^\text{ext} B \right)^2 - \Delta^2 \right]
\end{equation}
Thus, the area-splitting term $\alpha$ in Eq.~(\ref{eq:LK}) is given by:
\begin{equation}
\alpha = - \frac{\pi e\hbar \mu  \bm{g}_z^\text{ext} }{m_ev_x v_y},
\end{equation}
where we have used $\mu_B = e\hbar/2m_e$. Plugging in the result for $\bm{g}_z^\text{ext}$ from Eq.~(\ref{eq:gextfinal}),
\begin{equation}
\frac{\hbar\alpha}{e} =  - \frac{\pi \Delta }{\mu} - \frac{\pi \hbar^2 \mu }{m_ev_x v_y}\left[ \frac{g_z^p}{2}\left(1+\frac{\Delta}{\mu}\right)-\frac{g_z^s}{2}\left(1-\frac{\Delta}{\mu}\right) \right] \label{eq:alpha}
\end{equation}

Plugging in the calculation of the Berry phase (Eq.~(\ref{eq:berryphase})) and the area difference (Eq.~(\ref{eq:alpha})) to the Lifschitz-Kosevich formula (Eq.~(\ref{eq:LK})), we derive the expression for quantum oscillations:
\begin{widetext}
\begin{align}
    \Delta \rho 
                &\propto \cos\left(   \frac{\pi \hbar^2 \mu }{m_ev_x v_y}\left[ \frac{g_z^p}{2}\left(1+\frac{\Delta}{\mu}\right)-\frac{g_z^s}{2}\left(1-\frac{\Delta}{\mu}\right) \right]  \right)\cos\left(\frac{\hbar S_F}{eB}\pm\frac{\pi}{4}\right)
\end{align}
\end{widetext}

\section*{References}

\bibliographystyle{apsrev4-1}

\begin{thebibliography}{33}%
\makeatletter
\providecommand \@ifxundefined [1]{%
 \@ifx{#1\undefined}
}%
\providecommand \@ifnum [1]{%
 \ifnum #1\expandafter \@firstoftwo
 \else \expandafter \@secondoftwo
 \fi
}%
\providecommand \@ifx [1]{%
 \ifx #1\expandafter \@firstoftwo
 \else \expandafter \@secondoftwo
 \fi
}%
\providecommand \natexlab [1]{#1}%
\providecommand \enquote  [1]{``#1''}%
\providecommand \bibnamefont  [1]{#1}%
\providecommand \bibfnamefont [1]{#1}%
\providecommand \citenamefont [1]{#1}%
\providecommand \href@noop [0]{\@secondoftwo}%
\providecommand \href [0]{\begingroup \@sanitize@url \@href}%
\providecommand \@href[1]{\@@startlink{#1}\@@href}%
\providecommand \@@href[1]{\endgroup#1\@@endlink}%
\providecommand \@sanitize@url [0]{\catcode `\\12\catcode `\$12\catcode
  `\&12\catcode `\#12\catcode `\^12\catcode `\_12\catcode `\%12\relax}%
\providecommand \@@startlink[1]{}%
\providecommand \@@endlink[0]{}%
\providecommand \url  [0]{\begingroup\@sanitize@url \@url }%
\providecommand \@url [1]{\endgroup\@href {#1}{\urlprefix }}%
\providecommand \urlprefix  [0]{URL }%
\providecommand \Eprint [0]{\href }%
\providecommand \doibase [0]{http://dx.doi.org/}%
\providecommand \selectlanguage [0]{\@gobble}%
\providecommand \bibinfo  [0]{\@secondoftwo}%
\providecommand \bibfield  [0]{\@secondoftwo}%
\providecommand \translation [1]{[#1]}%
\providecommand \BibitemOpen [0]{}%
\providecommand \bibitemStop [0]{}%
\providecommand \bibitemNoStop [0]{.\EOS\space}%
\providecommand \EOS [0]{\spacefactor3000\relax}%
\providecommand \BibitemShut  [1]{\csname bibitem#1\endcsname}%
\let\auto@bib@innerbib\@empty
\bibitem [{\citenamefont {Kane}\ and\ \citenamefont
  {Mele}(2005{\natexlab{a}})}]{KaneMele2005}%
  \BibitemOpen
  \bibfield  {author} {\bibinfo {author} {\bibfnamefont {C.~L.}\ \bibnamefont
  {Kane}}\ and\ \bibinfo {author} {\bibfnamefont {E.~J.}\ \bibnamefont
  {Mele}},\ }\href {\doibase 10.1103/PhysRevLett.95.226801} {\bibfield
  {journal} {\bibinfo  {journal} {Phys. Rev. Lett.}\ }\textbf {\bibinfo
  {volume} {95}},\ \bibinfo {pages} {226801} (\bibinfo {year}
  {2005}{\natexlab{a}})}\BibitemShut {NoStop}%
\bibitem [{\citenamefont {Kane}\ and\ \citenamefont
  {Mele}(2005{\natexlab{b}})}]{KaneMele2005-2}%
  \BibitemOpen
  \bibfield  {author} {\bibinfo {author} {\bibfnamefont {C.~L.}\ \bibnamefont
  {Kane}}\ and\ \bibinfo {author} {\bibfnamefont {E.~J.}\ \bibnamefont
  {Mele}},\ }\href {\doibase 10.1103/PhysRevLett.95.146802} {\bibfield
  {journal} {\bibinfo  {journal} {Phys. Rev. Lett.}\ }\textbf {\bibinfo
  {volume} {95}},\ \bibinfo {pages} {146802} (\bibinfo {year}
  {2005}{\natexlab{b}})}\BibitemShut {NoStop}%
\bibitem [{\citenamefont {Bernevig}\ \emph {et~al.}(2006)\citenamefont
  {Bernevig}, \citenamefont {Hughes},\ and\ \citenamefont {Zhang}}]{BHZ2006}%
  \BibitemOpen
  \bibfield  {author} {\bibinfo {author} {\bibfnamefont {B.~A.}\ \bibnamefont
  {Bernevig}}, \bibinfo {author} {\bibfnamefont {T.~L.}\ \bibnamefont
  {Hughes}}, \ and\ \bibinfo {author} {\bibfnamefont {S.-C.}\ \bibnamefont
  {Zhang}},\ }\href {\doibase 10.1126/science.1133734} {\bibfield  {journal}
  {\bibinfo  {journal} {Science}\ }\textbf {\bibinfo {volume} {314}},\ \bibinfo
  {pages} {1757} (\bibinfo {year} {2006})},\ \Eprint
  {http://arxiv.org/abs/https://www.science.org/doi/pdf/10.1126/science.1133734}
  {https://www.science.org/doi/pdf/10.1126/science.1133734} \BibitemShut
  {NoStop}%
\bibitem [{\citenamefont {König}\ \emph {et~al.}(2007)\citenamefont {König},
  \citenamefont {Wiedmann}, \citenamefont {Brüne}, \citenamefont {Roth},
  \citenamefont {Buhmann}, \citenamefont {Molenkamp}, \citenamefont {Qi},\ and\
  \citenamefont {Zhang}}]{konig2007}%
  \BibitemOpen
  \bibfield  {author} {\bibinfo {author} {\bibfnamefont {M.}~\bibnamefont
  {König}}, \bibinfo {author} {\bibfnamefont {S.}~\bibnamefont {Wiedmann}},
  \bibinfo {author} {\bibfnamefont {C.}~\bibnamefont {Brüne}}, \bibinfo
  {author} {\bibfnamefont {A.}~\bibnamefont {Roth}}, \bibinfo {author}
  {\bibfnamefont {H.}~\bibnamefont {Buhmann}}, \bibinfo {author} {\bibfnamefont
  {L.~W.}\ \bibnamefont {Molenkamp}}, \bibinfo {author} {\bibfnamefont {X.-L.}\
  \bibnamefont {Qi}}, \ and\ \bibinfo {author} {\bibfnamefont {S.-C.}\
  \bibnamefont {Zhang}},\ }\href {\doibase 10.1126/science.1148047} {\bibfield
  {journal} {\bibinfo  {journal} {Science}\ }\textbf {\bibinfo {volume}
  {318}},\ \bibinfo {pages} {766} (\bibinfo {year} {2007})},\ \Eprint
  {http://arxiv.org/abs/https://www.science.org/doi/pdf/10.1126/science.1148047}
  {https://www.science.org/doi/pdf/10.1126/science.1148047} \BibitemShut
  {NoStop}%
\bibitem [{\citenamefont {Moore}\ and\ \citenamefont
  {Balents}(2007)}]{Moore2007}%
  \BibitemOpen
  \bibfield  {author} {\bibinfo {author} {\bibfnamefont {J.~E.}\ \bibnamefont
  {Moore}}\ and\ \bibinfo {author} {\bibfnamefont {L.}~\bibnamefont
  {Balents}},\ }\href {\doibase 10.1103/PhysRevB.75.121306} {\bibfield
  {journal} {\bibinfo  {journal} {Phys. Rev. B}\ }\textbf {\bibinfo {volume}
  {75}},\ \bibinfo {pages} {121306} (\bibinfo {year} {2007})}\BibitemShut
  {NoStop}%
\bibitem [{\citenamefont {Xu}\ \emph {et~al.}(2012)\citenamefont {Xu},
  \citenamefont {Liu}, \citenamefont {Alidoust}, \citenamefont {Neupane},
  \citenamefont {Qian}, \citenamefont {Belopolski}, \citenamefont {Denlinger},
  \citenamefont {Wang}, \citenamefont {Lin}, \citenamefont {Wray},
  \citenamefont {Landolt}, \citenamefont {Slomski}, \citenamefont {Dil},
  \citenamefont {Marcinkova}, \citenamefont {Morosan}, \citenamefont {Gibson},
  \citenamefont {Sankar}, \citenamefont {Chou}, \citenamefont {Cava},
  \citenamefont {Bansil},\ and\ \citenamefont {Hasan}}]{Xu2012}%
  \BibitemOpen
  \bibfield  {author} {\bibinfo {author} {\bibfnamefont {S.-Y.}\ \bibnamefont
  {Xu}}, \bibinfo {author} {\bibfnamefont {C.}~\bibnamefont {Liu}}, \bibinfo
  {author} {\bibfnamefont {N.}~\bibnamefont {Alidoust}}, \bibinfo {author}
  {\bibfnamefont {M.}~\bibnamefont {Neupane}}, \bibinfo {author} {\bibfnamefont
  {D.}~\bibnamefont {Qian}}, \bibinfo {author} {\bibfnamefont {I.}~\bibnamefont
  {Belopolski}}, \bibinfo {author} {\bibfnamefont {J.~D.}\ \bibnamefont
  {Denlinger}}, \bibinfo {author} {\bibfnamefont {Y.~J.}\ \bibnamefont {Wang}},
  \bibinfo {author} {\bibfnamefont {H.}~\bibnamefont {Lin}}, \bibinfo {author}
  {\bibfnamefont {L.~A.}\ \bibnamefont {Wray}}, \bibinfo {author}
  {\bibfnamefont {G.}~\bibnamefont {Landolt}}, \bibinfo {author} {\bibfnamefont
  {B.}~\bibnamefont {Slomski}}, \bibinfo {author} {\bibfnamefont {J.~H.}\
  \bibnamefont {Dil}}, \bibinfo {author} {\bibfnamefont {A.}~\bibnamefont
  {Marcinkova}}, \bibinfo {author} {\bibfnamefont {E.}~\bibnamefont {Morosan}},
  \bibinfo {author} {\bibfnamefont {Q.}~\bibnamefont {Gibson}}, \bibinfo
  {author} {\bibfnamefont {R.}~\bibnamefont {Sankar}}, \bibinfo {author}
  {\bibfnamefont {F.~C.}\ \bibnamefont {Chou}}, \bibinfo {author}
  {\bibfnamefont {R.~J.}\ \bibnamefont {Cava}}, \bibinfo {author}
  {\bibfnamefont {A.}~\bibnamefont {Bansil}}, \ and\ \bibinfo {author}
  {\bibfnamefont {M.~Z.}\ \bibnamefont {Hasan}},\ }\href {\doibase
  10.1038/ncomms2191} {\bibfield  {journal} {\bibinfo  {journal} {Nature
  Communications}\ }\textbf {\bibinfo {volume} {3}},\ \bibinfo {pages} {1192}
  (\bibinfo {year} {2012})}\BibitemShut {NoStop}%
\bibitem [{\citenamefont {Shoenberg}(1984)}]{shoenberg1984}%
  \BibitemOpen
  \bibfield  {author} {\bibinfo {author} {\bibfnamefont {D.}~\bibnamefont
  {Shoenberg}},\ }\href {\doibase 10.1017/CBO9780511897870} {\emph {\bibinfo
  {title} {Magnetic Oscillations in Metals}}},\ Cambridge Monographs on
  Physics\ (\bibinfo  {publisher} {Cambridge University Press},\ \bibinfo
  {year} {1984})\BibitemShut {NoStop}%
\bibitem [{\citenamefont {Mikitik}\ and\ \citenamefont
  {Sharlai}(1999)}]{Mikitik1999}%
  \BibitemOpen
  \bibfield  {author} {\bibinfo {author} {\bibfnamefont {G.~P.}\ \bibnamefont
  {Mikitik}}\ and\ \bibinfo {author} {\bibfnamefont {Y.~V.}\ \bibnamefont
  {Sharlai}},\ }\href {\doibase 10.1103/PhysRevLett.82.2147} {\bibfield
  {journal} {\bibinfo  {journal} {Phys. Rev. Lett.}\ }\textbf {\bibinfo
  {volume} {82}},\ \bibinfo {pages} {2147} (\bibinfo {year}
  {1999})}\BibitemShut {NoStop}%
\bibitem [{\citenamefont {Novoselov}\ \emph {et~al.}(2005)\citenamefont
  {Novoselov}, \citenamefont {Geim}, \citenamefont {Morozov}, \citenamefont
  {Jiang}, \citenamefont {Katsnelson}, \citenamefont {Grigorieva},
  \citenamefont {Dubonos},\ and\ \citenamefont {Firsov}}]{Novoselov2005}%
  \BibitemOpen
  \bibfield  {author} {\bibinfo {author} {\bibfnamefont {K.~S.}\ \bibnamefont
  {Novoselov}}, \bibinfo {author} {\bibfnamefont {A.~K.}\ \bibnamefont {Geim}},
  \bibinfo {author} {\bibfnamefont {S.~V.}\ \bibnamefont {Morozov}}, \bibinfo
  {author} {\bibfnamefont {D.}~\bibnamefont {Jiang}}, \bibinfo {author}
  {\bibfnamefont {M.~I.}\ \bibnamefont {Katsnelson}}, \bibinfo {author}
  {\bibfnamefont {I.~V.}\ \bibnamefont {Grigorieva}}, \bibinfo {author}
  {\bibfnamefont {S.~V.}\ \bibnamefont {Dubonos}}, \ and\ \bibinfo {author}
  {\bibfnamefont {A.~A.}\ \bibnamefont {Firsov}},\ }\href {\doibase
  10.1038/nature04233} {\bibfield  {journal} {\bibinfo  {journal} {Nature}\
  }\textbf {\bibinfo {volume} {438}},\ \bibinfo {pages} {197} (\bibinfo {year}
  {2005})}\BibitemShut {NoStop}%
\bibitem [{\citenamefont {Zhang}\ \emph {et~al.}(2005)\citenamefont {Zhang},
  \citenamefont {Tan}, \citenamefont {Stormer},\ and\ \citenamefont
  {Kim}}]{Zhang2005}%
  \BibitemOpen
  \bibfield  {author} {\bibinfo {author} {\bibfnamefont {Y.}~\bibnamefont
  {Zhang}}, \bibinfo {author} {\bibfnamefont {Y.-W.}\ \bibnamefont {Tan}},
  \bibinfo {author} {\bibfnamefont {H.~L.}\ \bibnamefont {Stormer}}, \ and\
  \bibinfo {author} {\bibfnamefont {P.}~\bibnamefont {Kim}},\ }\href {\doibase
  10.1038/nature04235} {\bibfield  {journal} {\bibinfo  {journal} {Nature}\
  }\textbf {\bibinfo {volume} {438}},\ \bibinfo {pages} {201} (\bibinfo {year}
  {2005})}\BibitemShut {NoStop}%
\bibitem [{\citenamefont {Luttinger}\ and\ \citenamefont
  {Kohn}(1955)}]{luttingerperturbation}%
  \BibitemOpen
  \bibfield  {author} {\bibinfo {author} {\bibfnamefont {J.~M.}\ \bibnamefont
  {Luttinger}}\ and\ \bibinfo {author} {\bibfnamefont {W.}~\bibnamefont
  {Kohn}},\ }\href {\doibase 10.1103/PhysRev.97.869} {\bibfield  {journal}
  {\bibinfo  {journal} {Physical Review}\ }\textbf {\bibinfo {volume} {97}},\
  \bibinfo {pages} {869} (\bibinfo {year} {1955})}\BibitemShut {NoStop}%
\bibitem [{\citenamefont {Cohen}\ and\ \citenamefont
  {Blount}(1960)}]{cohengfactor1960}%
  \BibitemOpen
  \bibfield  {author} {\bibinfo {author} {\bibfnamefont {M.~H.}\ \bibnamefont
  {Cohen}}\ and\ \bibinfo {author} {\bibfnamefont {E.~I.}\ \bibnamefont
  {Blount}},\ }\href {\doibase 10.1080/14786436008243294} {\bibfield  {journal}
  {\bibinfo  {journal} {Philosophical Magazine}\ }\textbf {\bibinfo {volume}
  {5}},\ \bibinfo {pages} {115} (\bibinfo {year} {1960})}\BibitemShut {NoStop}%
\bibitem [{\citenamefont {Wang}\ \emph {et~al.}(2018)\citenamefont {Wang},
  \citenamefont {Niu}, \citenamefont {Yan}, \citenamefont {Li}, \citenamefont
  {Bi}, \citenamefont {Yao}, \citenamefont {Yu},\ and\ \citenamefont
  {Wu}}]{Wang9145}%
  \BibitemOpen
  \bibfield  {author} {\bibinfo {author} {\bibfnamefont {J.}~\bibnamefont
  {Wang}}, \bibinfo {author} {\bibfnamefont {J.}~\bibnamefont {Niu}}, \bibinfo
  {author} {\bibfnamefont {B.}~\bibnamefont {Yan}}, \bibinfo {author}
  {\bibfnamefont {X.}~\bibnamefont {Li}}, \bibinfo {author} {\bibfnamefont
  {R.}~\bibnamefont {Bi}}, \bibinfo {author} {\bibfnamefont {Y.}~\bibnamefont
  {Yao}}, \bibinfo {author} {\bibfnamefont {D.}~\bibnamefont {Yu}}, \ and\
  \bibinfo {author} {\bibfnamefont {X.}~\bibnamefont {Wu}},\ }\href {\doibase
  10.1073/pnas.1804958115} {\bibfield  {journal} {\bibinfo  {journal}
  {Proceedings of the National Academy of Sciences}\ }\textbf {\bibinfo
  {volume} {115}},\ \bibinfo {pages} {9145} (\bibinfo {year} {2018})},\ \Eprint
  {http://arxiv.org/abs/https://www.pnas.org/content/115/37/9145.full.pdf}
  {https://www.pnas.org/content/115/37/9145.full.pdf} \BibitemShut {NoStop}%
\bibitem [{\citenamefont {Sun}\ \emph {et~al.}(2020)\citenamefont {Sun},
  \citenamefont {Song}, \citenamefont {Weng},\ and\ \citenamefont
  {Dai}}]{PhysRevB.101.125118}%
  \BibitemOpen
  \bibfield  {author} {\bibinfo {author} {\bibfnamefont {S.}~\bibnamefont
  {Sun}}, \bibinfo {author} {\bibfnamefont {Z.}~\bibnamefont {Song}}, \bibinfo
  {author} {\bibfnamefont {H.}~\bibnamefont {Weng}}, \ and\ \bibinfo {author}
  {\bibfnamefont {X.}~\bibnamefont {Dai}},\ }\href {\doibase
  10.1103/PhysRevB.101.125118} {\bibfield  {journal} {\bibinfo  {journal}
  {Physical Review B}\ }\textbf {\bibinfo {volume} {101}},\ \bibinfo {pages}
  {125118} (\bibinfo {year} {2020})}\BibitemShut {NoStop}%
\bibitem [{\citenamefont {Zhang}\ \emph {et~al.}(2017)\citenamefont {Zhang},
  \citenamefont {Wang}, \citenamefont {Yu}, \citenamefont {Liu}, \citenamefont
  {Liang}, \citenamefont {Huang}, \citenamefont {Nie}, \citenamefont {Sun},
  \citenamefont {Zhang}, \citenamefont {Shen},\ and\ \citenamefont
  {et~al.}}]{NatComm.15512}%
  \BibitemOpen
  \bibfield  {author} {\bibinfo {author} {\bibfnamefont {Y.}~\bibnamefont
  {Zhang}}, \bibinfo {author} {\bibfnamefont {C.}~\bibnamefont {Wang}},
  \bibinfo {author} {\bibfnamefont {L.}~\bibnamefont {Yu}}, \bibinfo {author}
  {\bibfnamefont {G.}~\bibnamefont {Liu}}, \bibinfo {author} {\bibfnamefont
  {A.}~\bibnamefont {Liang}}, \bibinfo {author} {\bibfnamefont
  {J.}~\bibnamefont {Huang}}, \bibinfo {author} {\bibfnamefont
  {S.}~\bibnamefont {Nie}}, \bibinfo {author} {\bibfnamefont {X.}~\bibnamefont
  {Sun}}, \bibinfo {author} {\bibfnamefont {Y.}~\bibnamefont {Zhang}}, \bibinfo
  {author} {\bibfnamefont {B.}~\bibnamefont {Shen}}, \ and\ \bibinfo {author}
  {\bibnamefont {et~al.}},\ }\href
  {https://www.nature.com/articles/ncomms15512} {\bibfield  {journal} {\bibinfo
   {journal} {Nature Communications}\ }\textbf {\bibinfo {volume} {8}}
  (\bibinfo {year} {2017})}\BibitemShut {NoStop}%
\bibitem [{\citenamefont {Chi}\ \emph {et~al.}(2017)\citenamefont {Chi},
  \citenamefont {Zhang}, \citenamefont {Gu}, \citenamefont {Kharzeev},
  \citenamefont {Dai},\ and\ \citenamefont {Li}}]{Chi2017}%
  \BibitemOpen
  \bibfield  {author} {\bibinfo {author} {\bibfnamefont {H.}~\bibnamefont
  {Chi}}, \bibinfo {author} {\bibfnamefont {C.}~\bibnamefont {Zhang}}, \bibinfo
  {author} {\bibfnamefont {G.}~\bibnamefont {Gu}}, \bibinfo {author}
  {\bibfnamefont {D.~E.}\ \bibnamefont {Kharzeev}}, \bibinfo {author}
  {\bibfnamefont {X.}~\bibnamefont {Dai}}, \ and\ \bibinfo {author}
  {\bibfnamefont {Q.}~\bibnamefont {Li}},\ }\href {\doibase
  10.1088/1367-2630/aa55a3} {\bibfield  {journal} {\bibinfo  {journal} {New
  Journal of Physics}\ }\textbf {\bibinfo {volume} {19}},\ \bibinfo {pages}
  {015005} (\bibinfo {year} {2017})}\BibitemShut {NoStop}%
\bibitem [{\citenamefont {Li}\ \emph {et~al.}(2016)\citenamefont {Li},
  \citenamefont {Kharzeev}, \citenamefont {Zhang}, \citenamefont {Huang},
  \citenamefont {Pletikosi{\'{c}}}, \citenamefont {Fedorov}, \citenamefont
  {Zhong}, \citenamefont {Schneeloch}, \citenamefont {Gu},\ and\ \citenamefont
  {Valla}}]{Li2016}%
  \BibitemOpen
  \bibfield  {author} {\bibinfo {author} {\bibfnamefont {Q.}~\bibnamefont
  {Li}}, \bibinfo {author} {\bibfnamefont {D.~E.}\ \bibnamefont {Kharzeev}},
  \bibinfo {author} {\bibfnamefont {C.}~\bibnamefont {Zhang}}, \bibinfo
  {author} {\bibfnamefont {Y.}~\bibnamefont {Huang}}, \bibinfo {author}
  {\bibfnamefont {I.}~\bibnamefont {Pletikosi{\'{c}}}}, \bibinfo {author}
  {\bibfnamefont {A.~V.}\ \bibnamefont {Fedorov}}, \bibinfo {author}
  {\bibfnamefont {R.~D.}\ \bibnamefont {Zhong}}, \bibinfo {author}
  {\bibfnamefont {J.~A.}\ \bibnamefont {Schneeloch}}, \bibinfo {author}
  {\bibfnamefont {G.~D.}\ \bibnamefont {Gu}}, \ and\ \bibinfo {author}
  {\bibfnamefont {T.}~\bibnamefont {Valla}},\ }\href {\doibase
  10.1038/nphys3648} {\bibfield  {journal} {\bibinfo  {journal} {Nature
  Physics}\ }\textbf {\bibinfo {volume} {12}},\ \bibinfo {pages} {550}
  (\bibinfo {year} {2016})}\BibitemShut {NoStop}%
\bibitem [{\citenamefont {Tang}\ \emph {et~al.}(2019)\citenamefont {Tang},
  \citenamefont {Ren}, \citenamefont {Wang}, \citenamefont {Zhong},
  \citenamefont {Schneeloch}, \citenamefont {Yang}, \citenamefont {Yang},
  \citenamefont {Lee}, \citenamefont {Gu}, \citenamefont {Qiao},\ and\
  \citenamefont {Zhang}}]{Tang2019}%
  \BibitemOpen
  \bibfield  {author} {\bibinfo {author} {\bibfnamefont {F.}~\bibnamefont
  {Tang}}, \bibinfo {author} {\bibfnamefont {Y.}~\bibnamefont {Ren}}, \bibinfo
  {author} {\bibfnamefont {P.}~\bibnamefont {Wang}}, \bibinfo {author}
  {\bibfnamefont {R.}~\bibnamefont {Zhong}}, \bibinfo {author} {\bibfnamefont
  {J.}~\bibnamefont {Schneeloch}}, \bibinfo {author} {\bibfnamefont {S.~A.}\
  \bibnamefont {Yang}}, \bibinfo {author} {\bibfnamefont {K.}~\bibnamefont
  {Yang}}, \bibinfo {author} {\bibfnamefont {P.~A.}\ \bibnamefont {Lee}},
  \bibinfo {author} {\bibfnamefont {G.}~\bibnamefont {Gu}}, \bibinfo {author}
  {\bibfnamefont {Z.}~\bibnamefont {Qiao}}, \ and\ \bibinfo {author}
  {\bibfnamefont {L.}~\bibnamefont {Zhang}},\ }\href {\doibase
  10.1038/s41586-019-1180-9} {\bibfield  {journal} {\bibinfo  {journal}
  {Nature}\ }\textbf {\bibinfo {volume} {569}},\ \bibinfo {pages} {537}
  (\bibinfo {year} {2019})}\BibitemShut {NoStop}%
\bibitem [{\citenamefont {Liu}\ \emph {et~al.}(2016)\citenamefont {Liu},
  \citenamefont {Yuan}, \citenamefont {Zhang}, \citenamefont {Jin},
  \citenamefont {Narayan}, \citenamefont {Luo}, \citenamefont {Chen},
  \citenamefont {Yang}, \citenamefont {Zou}, \citenamefont {Wu}, \citenamefont
  {Sanvito}, \citenamefont {Xia}, \citenamefont {Li}, \citenamefont {Wang},\
  and\ \citenamefont {Xiu}}]{Liu2016}%
  \BibitemOpen
  \bibfield  {author} {\bibinfo {author} {\bibfnamefont {Y.}~\bibnamefont
  {Liu}}, \bibinfo {author} {\bibfnamefont {X.}~\bibnamefont {Yuan}}, \bibinfo
  {author} {\bibfnamefont {C.}~\bibnamefont {Zhang}}, \bibinfo {author}
  {\bibfnamefont {Z.}~\bibnamefont {Jin}}, \bibinfo {author} {\bibfnamefont
  {A.}~\bibnamefont {Narayan}}, \bibinfo {author} {\bibfnamefont
  {C.}~\bibnamefont {Luo}}, \bibinfo {author} {\bibfnamefont {Z.}~\bibnamefont
  {Chen}}, \bibinfo {author} {\bibfnamefont {L.}~\bibnamefont {Yang}}, \bibinfo
  {author} {\bibfnamefont {J.}~\bibnamefont {Zou}}, \bibinfo {author}
  {\bibfnamefont {X.}~\bibnamefont {Wu}}, \bibinfo {author} {\bibfnamefont
  {S.}~\bibnamefont {Sanvito}}, \bibinfo {author} {\bibfnamefont
  {Z.}~\bibnamefont {Xia}}, \bibinfo {author} {\bibfnamefont {L.}~\bibnamefont
  {Li}}, \bibinfo {author} {\bibfnamefont {Z.}~\bibnamefont {Wang}}, \ and\
  \bibinfo {author} {\bibfnamefont {F.}~\bibnamefont {Xiu}},\ }\href {\doibase
  10.1038/ncomms12516} {\bibfield  {journal} {\bibinfo  {journal} {Nature
  Communications}\ }\textbf {\bibinfo {volume} {7}},\ \bibinfo {pages} {12516}
  (\bibinfo {year} {2016})}\BibitemShut {NoStop}%
\bibitem [{\citenamefont {Chen}\ \emph {et~al.}(2017)\citenamefont {Chen},
  \citenamefont {Chen}, \citenamefont {Zhong}, \citenamefont {Schneeloch},
  \citenamefont {Zhang}, \citenamefont {Huang}, \citenamefont {Qu},
  \citenamefont {Yu}, \citenamefont {Li}, \citenamefont {Gu},\ and\
  \citenamefont {Wang}}]{Chen816}%
  \BibitemOpen
  \bibfield  {author} {\bibinfo {author} {\bibfnamefont {Z.-G.}\ \bibnamefont
  {Chen}}, \bibinfo {author} {\bibfnamefont {R.~Y.}\ \bibnamefont {Chen}},
  \bibinfo {author} {\bibfnamefont {R.~D.}\ \bibnamefont {Zhong}}, \bibinfo
  {author} {\bibfnamefont {J.}~\bibnamefont {Schneeloch}}, \bibinfo {author}
  {\bibfnamefont {C.}~\bibnamefont {Zhang}}, \bibinfo {author} {\bibfnamefont
  {Y.}~\bibnamefont {Huang}}, \bibinfo {author} {\bibfnamefont
  {F.}~\bibnamefont {Qu}}, \bibinfo {author} {\bibfnamefont {R.}~\bibnamefont
  {Yu}}, \bibinfo {author} {\bibfnamefont {Q.}~\bibnamefont {Li}}, \bibinfo
  {author} {\bibfnamefont {G.~D.}\ \bibnamefont {Gu}}, \ and\ \bibinfo {author}
  {\bibfnamefont {N.~L.}\ \bibnamefont {Wang}},\ }\href {\doibase
  10.1073/pnas.1613110114} {\bibfield  {journal} {\bibinfo  {journal}
  {Proceedings of the National Academy of Sciences}\ }\textbf {\bibinfo
  {volume} {114}},\ \bibinfo {pages} {816} (\bibinfo {year} {2017})},\ \Eprint
  {http://arxiv.org/abs/https://www.pnas.org/content/114/5/816.full.pdf}
  {https://www.pnas.org/content/114/5/816.full.pdf} \BibitemShut {NoStop}%
\bibitem [{\citenamefont {Jiang}\ \emph {et~al.}(2017)\citenamefont {Jiang},
  \citenamefont {Dun}, \citenamefont {Zhou}, \citenamefont {Lu}, \citenamefont
  {Chen}, \citenamefont {Moon}, \citenamefont {Besara}, \citenamefont
  {Siegrist}, \citenamefont {Baumbach}, \citenamefont {Smirnov},\ and\
  \citenamefont {Jiang}}]{PhysRevB.96.041101}%
  \BibitemOpen
  \bibfield  {author} {\bibinfo {author} {\bibfnamefont {Y.}~\bibnamefont
  {Jiang}}, \bibinfo {author} {\bibfnamefont {Z.~L.}\ \bibnamefont {Dun}},
  \bibinfo {author} {\bibfnamefont {H.~D.}\ \bibnamefont {Zhou}}, \bibinfo
  {author} {\bibfnamefont {Z.}~\bibnamefont {Lu}}, \bibinfo {author}
  {\bibfnamefont {K.-W.}\ \bibnamefont {Chen}}, \bibinfo {author}
  {\bibfnamefont {S.}~\bibnamefont {Moon}}, \bibinfo {author} {\bibfnamefont
  {T.}~\bibnamefont {Besara}}, \bibinfo {author} {\bibfnamefont {T.~M.}\
  \bibnamefont {Siegrist}}, \bibinfo {author} {\bibfnamefont {R.~E.}\
  \bibnamefont {Baumbach}}, \bibinfo {author} {\bibfnamefont {D.}~\bibnamefont
  {Smirnov}}, \ and\ \bibinfo {author} {\bibfnamefont {Z.}~\bibnamefont
  {Jiang}},\ }\href {\doibase 10.1103/PhysRevB.96.041101} {\bibfield  {journal}
  {\bibinfo  {journal} {Phys. Rev. B}\ }\textbf {\bibinfo {volume} {96}},\
  \bibinfo {pages} {041101} (\bibinfo {year} {2017})}\BibitemShut {NoStop}%
\bibitem [{\citenamefont {Xu}\ \emph {et~al.}(2018)\citenamefont {Xu},
  \citenamefont {Zhao}, \citenamefont {Marsik}, \citenamefont {Sheveleva},
  \citenamefont {Lyzwa}, \citenamefont {Dai}, \citenamefont {Chen},
  \citenamefont {Qiu},\ and\ \citenamefont
  {Bernhard}}]{PhysRevLett.121.187401}%
  \BibitemOpen
  \bibfield  {author} {\bibinfo {author} {\bibfnamefont {B.}~\bibnamefont
  {Xu}}, \bibinfo {author} {\bibfnamefont {L.~X.}\ \bibnamefont {Zhao}},
  \bibinfo {author} {\bibfnamefont {P.}~\bibnamefont {Marsik}}, \bibinfo
  {author} {\bibfnamefont {E.}~\bibnamefont {Sheveleva}}, \bibinfo {author}
  {\bibfnamefont {F.}~\bibnamefont {Lyzwa}}, \bibinfo {author} {\bibfnamefont
  {Y.~M.}\ \bibnamefont {Dai}}, \bibinfo {author} {\bibfnamefont {G.~F.}\
  \bibnamefont {Chen}}, \bibinfo {author} {\bibfnamefont {X.~G.}\ \bibnamefont
  {Qiu}}, \ and\ \bibinfo {author} {\bibfnamefont {C.}~\bibnamefont
  {Bernhard}},\ }\href {\doibase 10.1103/PhysRevLett.121.187401} {\bibfield
  {journal} {\bibinfo  {journal} {Phys. Rev. Lett.}\ }\textbf {\bibinfo
  {volume} {121}},\ \bibinfo {pages} {187401} (\bibinfo {year}
  {2018})}\BibitemShut {NoStop}%
\bibitem [{\citenamefont {Xiong}\ \emph {et~al.}(2017)\citenamefont {Xiong},
  \citenamefont {Sobota}, \citenamefont {Yang}, \citenamefont {Soifer},
  \citenamefont {Gauthier}, \citenamefont {Lu}, \citenamefont {Lv},
  \citenamefont {Yao}, \citenamefont {Lu}, \citenamefont {Hashimoto},
  \citenamefont {Kirchmann}, \citenamefont {Chen},\ and\ \citenamefont
  {Shen}}]{PhysRevB.95.195119}%
  \BibitemOpen
  \bibfield  {author} {\bibinfo {author} {\bibfnamefont {H.}~\bibnamefont
  {Xiong}}, \bibinfo {author} {\bibfnamefont {J.~A.}\ \bibnamefont {Sobota}},
  \bibinfo {author} {\bibfnamefont {S.-L.}\ \bibnamefont {Yang}}, \bibinfo
  {author} {\bibfnamefont {H.}~\bibnamefont {Soifer}}, \bibinfo {author}
  {\bibfnamefont {A.}~\bibnamefont {Gauthier}}, \bibinfo {author}
  {\bibfnamefont {M.-H.}\ \bibnamefont {Lu}}, \bibinfo {author} {\bibfnamefont
  {Y.-Y.}\ \bibnamefont {Lv}}, \bibinfo {author} {\bibfnamefont {S.-H.}\
  \bibnamefont {Yao}}, \bibinfo {author} {\bibfnamefont {D.}~\bibnamefont
  {Lu}}, \bibinfo {author} {\bibfnamefont {M.}~\bibnamefont {Hashimoto}},
  \bibinfo {author} {\bibfnamefont {P.~S.}\ \bibnamefont {Kirchmann}}, \bibinfo
  {author} {\bibfnamefont {Y.-F.}\ \bibnamefont {Chen}}, \ and\ \bibinfo
  {author} {\bibfnamefont {Z.-X.}\ \bibnamefont {Shen}},\ }\href {\doibase
  10.1103/PhysRevB.95.195119} {\bibfield  {journal} {\bibinfo  {journal} {Phys.
  Rev. B}\ }\textbf {\bibinfo {volume} {95}},\ \bibinfo {pages} {195119}
  (\bibinfo {year} {2017})}\BibitemShut {NoStop}%
\bibitem [{\citenamefont {Weng}\ \emph {et~al.}(2014)\citenamefont {Weng},
  \citenamefont {Dai},\ and\ \citenamefont {Fang}}]{PhysRevX.4.011002}%
  \BibitemOpen
  \bibfield  {author} {\bibinfo {author} {\bibfnamefont {H.}~\bibnamefont
  {Weng}}, \bibinfo {author} {\bibfnamefont {X.}~\bibnamefont {Dai}}, \ and\
  \bibinfo {author} {\bibfnamefont {Z.}~\bibnamefont {Fang}},\ }\href {\doibase
  10.1103/PhysRevX.4.011002} {\bibfield  {journal} {\bibinfo  {journal} {Phys.
  Rev. X}\ }\textbf {\bibinfo {volume} {4}},\ \bibinfo {pages} {011002}
  (\bibinfo {year} {2014})}\BibitemShut {NoStop}%
\bibitem [{\citenamefont {Fan}\ \emph {et~al.}(2017)\citenamefont {Fan},
  \citenamefont {Liang}, \citenamefont {Chen}, \citenamefont {Yao},\ and\
  \citenamefont {Zhou}}]{SciRep.7.45667}%
  \BibitemOpen
  \bibfield  {author} {\bibinfo {author} {\bibfnamefont {Z.}~\bibnamefont
  {Fan}}, \bibinfo {author} {\bibfnamefont {Q.-F.}\ \bibnamefont {Liang}},
  \bibinfo {author} {\bibfnamefont {Y.~B.}\ \bibnamefont {Chen}}, \bibinfo
  {author} {\bibfnamefont {S.-H.}\ \bibnamefont {Yao}}, \ and\ \bibinfo
  {author} {\bibfnamefont {J.}~\bibnamefont {Zhou}},\ }\href@noop {} {\bibfield
   {journal} {\bibinfo  {journal} {Scientific Reports}\ }\textbf {\bibinfo
  {volume} {7}} (\bibinfo {year} {2017})}\BibitemShut {NoStop}%
\bibitem [{\citenamefont {Zhang}\ \emph {et~al.}(2021)\citenamefont {Zhang},
  \citenamefont {Noguchi}, \citenamefont {Kuroda}, \citenamefont {Lin},
  \citenamefont {Kawaguchi}, \citenamefont {Yaji}, \citenamefont {Harasawa},
  \citenamefont {Lippmaa}, \citenamefont {Nie}, \citenamefont {Weng},
  \citenamefont {Kandyba}, \citenamefont {Giampietri}, \citenamefont {Barinov},
  \citenamefont {Li}, \citenamefont {Gu}, \citenamefont {Shin},\ and\
  \citenamefont {Kondo}}]{Zhang2021}%
  \BibitemOpen
  \bibfield  {author} {\bibinfo {author} {\bibfnamefont {P.}~\bibnamefont
  {Zhang}}, \bibinfo {author} {\bibfnamefont {R.}~\bibnamefont {Noguchi}},
  \bibinfo {author} {\bibfnamefont {K.}~\bibnamefont {Kuroda}}, \bibinfo
  {author} {\bibfnamefont {C.}~\bibnamefont {Lin}}, \bibinfo {author}
  {\bibfnamefont {K.}~\bibnamefont {Kawaguchi}}, \bibinfo {author}
  {\bibfnamefont {K.}~\bibnamefont {Yaji}}, \bibinfo {author} {\bibfnamefont
  {A.}~\bibnamefont {Harasawa}}, \bibinfo {author} {\bibfnamefont
  {M.}~\bibnamefont {Lippmaa}}, \bibinfo {author} {\bibfnamefont
  {S.}~\bibnamefont {Nie}}, \bibinfo {author} {\bibfnamefont {H.}~\bibnamefont
  {Weng}}, \bibinfo {author} {\bibfnamefont {V.}~\bibnamefont {Kandyba}},
  \bibinfo {author} {\bibfnamefont {A.}~\bibnamefont {Giampietri}}, \bibinfo
  {author} {\bibfnamefont {A.}~\bibnamefont {Barinov}}, \bibinfo {author}
  {\bibfnamefont {Q.}~\bibnamefont {Li}}, \bibinfo {author} {\bibfnamefont
  {G.~D.}\ \bibnamefont {Gu}}, \bibinfo {author} {\bibfnamefont
  {S.}~\bibnamefont {Shin}}, \ and\ \bibinfo {author} {\bibfnamefont
  {T.}~\bibnamefont {Kondo}},\ }\href {\doibase 10.1038/s41467-020-20564-8}
  {\bibfield  {journal} {\bibinfo  {journal} {Nature Communications}\ }\textbf
  {\bibinfo {volume} {12}},\ \bibinfo {pages} {406} (\bibinfo {year}
  {2021})}\BibitemShut {NoStop}%
\bibitem [{\citenamefont {Mutch}\ \emph {et~al.}(2019)\citenamefont {Mutch},
  \citenamefont {Chen}, \citenamefont {Went}, \citenamefont {Qian},
  \citenamefont {Wilson}, \citenamefont {Andreev}, \citenamefont {Chen},\ and\
  \citenamefont {Chu}}]{Mutcheaav9771}%
  \BibitemOpen
  \bibfield  {author} {\bibinfo {author} {\bibfnamefont {J.}~\bibnamefont
  {Mutch}}, \bibinfo {author} {\bibfnamefont {W.-C.}\ \bibnamefont {Chen}},
  \bibinfo {author} {\bibfnamefont {P.}~\bibnamefont {Went}}, \bibinfo {author}
  {\bibfnamefont {T.}~\bibnamefont {Qian}}, \bibinfo {author} {\bibfnamefont
  {I.~Z.}\ \bibnamefont {Wilson}}, \bibinfo {author} {\bibfnamefont
  {A.}~\bibnamefont {Andreev}}, \bibinfo {author} {\bibfnamefont {C.-C.}\
  \bibnamefont {Chen}}, \ and\ \bibinfo {author} {\bibfnamefont {J.-H.}\
  \bibnamefont {Chu}},\ }\href {\doibase 10.1126/sciadv.aav9771} {\bibfield
  {journal} {\bibinfo  {journal} {Science Advances}\ }\textbf {\bibinfo
  {volume} {5}} (\bibinfo {year} {2019}),\ 10.1126/sciadv.aav9771},\ \Eprint
  {http://arxiv.org/abs/https://advances.sciencemag.org/content/5/8/eaav9771.full.pdf}
  {https://advances.sciencemag.org/content/5/8/eaav9771.full.pdf} \BibitemShut
  {NoStop}%
\bibitem [{\citenamefont {Gusynin}\ and\ \citenamefont
  {Sharapov}(2005)}]{PhysRevB.71.125124}%
  \BibitemOpen
  \bibfield  {author} {\bibinfo {author} {\bibfnamefont {V.~P.}\ \bibnamefont
  {Gusynin}}\ and\ \bibinfo {author} {\bibfnamefont {S.~G.}\ \bibnamefont
  {Sharapov}},\ }\href {\doibase 10.1103/PhysRevB.71.125124} {\bibfield
  {journal} {\bibinfo  {journal} {Phys. Rev. B}\ }\textbf {\bibinfo {volume}
  {71}},\ \bibinfo {pages} {125124} (\bibinfo {year} {2005})}\BibitemShut
  {NoStop}%
\bibitem [{\citenamefont {Song}\ \emph {et~al.}(2021)\citenamefont {Song},
  \citenamefont {Sun}, \citenamefont {Xu}, \citenamefont {Nie}, \citenamefont
  {Weng}, \citenamefont {Fang},\ and\ \citenamefont {Dai}}]{song2021first}%
  \BibitemOpen
  \bibfield  {author} {\bibinfo {author} {\bibfnamefont {Z.}~\bibnamefont
  {Song}}, \bibinfo {author} {\bibfnamefont {S.}~\bibnamefont {Sun}}, \bibinfo
  {author} {\bibfnamefont {Y.}~\bibnamefont {Xu}}, \bibinfo {author}
  {\bibfnamefont {S.}~\bibnamefont {Nie}}, \bibinfo {author} {\bibfnamefont
  {H.}~\bibnamefont {Weng}}, \bibinfo {author} {\bibfnamefont {Z.}~\bibnamefont
  {Fang}}, \ and\ \bibinfo {author} {\bibfnamefont {X.}~\bibnamefont {Dai}},\
  }\enquote {\bibinfo {title} {First principle calculation of the effective
  zeeman's couplings in topological materials},}\ in\ \href {\doibase
  10.1142/9789811231711_0013} {\emph {\bibinfo {booktitle} {Memorial Volume for
  Shoucheng Zhang}}}\ (\bibinfo  {publisher} {World Scientific},\ \bibinfo
  {year} {2021})\ Chap.\ \bibinfo {chapter} {Chapter 11}, pp.\ \bibinfo {pages}
  {263--281},\ \Eprint
  {http://arxiv.org/abs/https://www.worldscientific.com/doi/pdf/10.1142/9789811231711_0013}
  {https://www.worldscientific.com/doi/pdf/10.1142/9789811231711_0013}
  \BibitemShut {NoStop}%
\bibitem [{\citenamefont {Mills}\ \emph {et~al.}(2019)\citenamefont {Mills},
  \citenamefont {Mizuno}, \citenamefont {Wang}, \citenamefont {Lyu},
  \citenamefont {Watanabe}, \citenamefont {Taniguchi}, \citenamefont {Camino},
  \citenamefont {Zhang},\ and\ \citenamefont {Du}}]{Mills2019}%
  \BibitemOpen
  \bibfield  {author} {\bibinfo {author} {\bibfnamefont {S.}~\bibnamefont
  {Mills}}, \bibinfo {author} {\bibfnamefont {N.}~\bibnamefont {Mizuno}},
  \bibinfo {author} {\bibfnamefont {P.}~\bibnamefont {Wang}}, \bibinfo {author}
  {\bibfnamefont {J.}~\bibnamefont {Lyu}}, \bibinfo {author} {\bibfnamefont
  {K.}~\bibnamefont {Watanabe}}, \bibinfo {author} {\bibfnamefont
  {T.}~\bibnamefont {Taniguchi}}, \bibinfo {author} {\bibfnamefont
  {F.}~\bibnamefont {Camino}}, \bibinfo {author} {\bibfnamefont
  {L.}~\bibnamefont {Zhang}}, \ and\ \bibinfo {author} {\bibfnamefont
  {X.}~\bibnamefont {Du}},\ }\href {\doibase 10.1088/2515-7639/ab1863}
  {\bibfield  {journal} {\bibinfo  {journal} {Journal of Physics: Materials}\
  }\textbf {\bibinfo {volume} {2}},\ \bibinfo {pages} {035003} (\bibinfo {year}
  {2019})}\BibitemShut {NoStop}%
\bibitem [{\citenamefont {Guan}\ \emph {et~al.}(2015)\citenamefont {Guan},
  \citenamefont {Kumaravadivel}, \citenamefont {Averin},\ and\ \citenamefont
  {Du}}]{Guan107.193102}%
  \BibitemOpen
  \bibfield  {author} {\bibinfo {author} {\bibfnamefont {F.}~\bibnamefont
  {Guan}}, \bibinfo {author} {\bibfnamefont {P.}~\bibnamefont {Kumaravadivel}},
  \bibinfo {author} {\bibfnamefont {D.~V.}\ \bibnamefont {Averin}}, \ and\
  \bibinfo {author} {\bibfnamefont {X.}~\bibnamefont {Du}},\ }\href {\doibase
  10.1063/1.4935239} {\bibfield  {journal} {\bibinfo  {journal} {Applied
  Physics Letters}\ }\textbf {\bibinfo {volume} {107}},\ \bibinfo {pages}
  {193102} (\bibinfo {year} {2015})},\ \Eprint
  {http://arxiv.org/abs/https://doi.org/10.1063/1.4935239}
  {https://doi.org/10.1063/1.4935239} \BibitemShut {NoStop}%
\bibitem [{\citenamefont {Guan}\ and\ \citenamefont {Du}(2017)}]{Guan2017}%
  \BibitemOpen
  \bibfield  {author} {\bibinfo {author} {\bibfnamefont {F.}~\bibnamefont
  {Guan}}\ and\ \bibinfo {author} {\bibfnamefont {X.}~\bibnamefont {Du}},\
  }\href {\doibase 10.1021/acs.nanolett.7b03618} {\bibfield  {journal}
  {\bibinfo  {journal} {Nano Letters}\ }\textbf {\bibinfo {volume} {17}},\
  \bibinfo {pages} {7009} (\bibinfo {year} {2017})}\BibitemShut {NoStop}%
\bibitem [{\citenamefont {Xiao}\ \emph {et~al.}(2010)\citenamefont {Xiao},
  \citenamefont {Chang},\ and\ \citenamefont {Niu}}]{Xiao2010Berry}%
  \BibitemOpen
  \bibfield  {author} {\bibinfo {author} {\bibfnamefont {D.}~\bibnamefont
  {Xiao}}, \bibinfo {author} {\bibfnamefont {M.-C.}\ \bibnamefont {Chang}}, \
  and\ \bibinfo {author} {\bibfnamefont {Q.}~\bibnamefont {Niu}},\ }\href
  {\doibase 10.1103/RevModPhys.82.1959} {\bibfield  {journal} {\bibinfo
  {journal} {Rev. Mod. Phys.}\ }\textbf {\bibinfo {volume} {82}},\ \bibinfo
  {pages} {1959} (\bibinfo {year} {2010})}\BibitemShut {NoStop}%
\end{thebibliography}

%

\section*{Acknowledgement}
X.D. acknowledges support from the National Science Foundation (NSF) under award DMR-1808491. X.D. and J.C. thank Aris Alexandradinata for insightful discussions.
J.C. acknowledges the support of the Flatiron Institute, a division of Simons Foundation, and support from the National Science Foundation under Grant No. DMR-1942447.

\end{document}